\documentstyle[eqsecnum,aps]{revtex}
\begin{document}
\title{Two-phonon background for the double giant resonance}
\author{V.Yu. Ponomarev$^{a}$ and C.A. Bertulani$^{b}$}
\address{$^a$Bogoliubov Laboratory of Theoretical\\
Physics, Joint Institute for Nuclear Research, 141980, Dubna, Russia}
\address{$^b$Instituto de F\'\i sica,\ 
Universidade Federal do Rio de Janeiro\\
21945-970 Rio de Janeiro, RJ, Brazil,\ 
E-mail: bertu@if.ufrj.br}
\date{\today}
\maketitle

\begin{abstract}
\end{abstract}

%\draft
%\preprint{Draft}

\bigskip


\bigskip

%\narrowtext
\twocolumn

One of the most exciting features of the recently discovered double giant
dipole resonance (DGDR) is the absolute value of its excitation cross
section in relativistic heavy ion collisions. These values were extracted
from total cross sections by separating a contribution arising from the
excitation of single giant dipole (GDR) and quadrupole (GQR) resonances and
were found to be enhanced by factors of order of 2-3 for $^{136}$Xe \cite
{sch93} and $^{197}$Au \cite{aum93} as compared to any theoretical
calculations available. Although for $^{208}$Pb \cite{bor96} the
experiment-theory correspondence is much better, theoretical calculations
still underpredict the DGDR cross section by about 30\%.

At the present time we have an experimental proof \cite{bor96} that Coulomb
excitation of the DGDR in relativistic heavy ion collisions occurs in a
two-step process. % via the GDR as an intermediate state.
The DGDR %or $[\mbox{GDR} \otimes \mbox{GDR}]$
states are embedded in a sea of other two-phonon states. In this paper we
consider the sea contribution to the total cross sections. It will be
concluded that numerous two-phonon states, other than $[\mbox{GDR} \otimes %
\mbox{GDR}]$ ones, excited in one-step processes form a physical background
in the DGDR energy region which has to be taken into account in the analysis
of experimental data.

The first investigation of the direct excitation of two-phonon states at
high energies from the ground state was done in refs.~\cite{cat92,pon92}.
The main purpose of these papers was to look for resonance structures in
photoexcitation %(and heavy ion collision at intermediate energies)
cross sections related to the excitation of rather selected two-phonon
configurations. 
%like $[\mbox{GDR} \otimes \mbox{GDR}]$, $[\mbox{GDR} \otimes
%\mbox{GQR}]$ and $[\mbox{GQR} \otimes \mbox{GQR}]$.
Although in ref.~\cite{pon92} it was pointed out that $[\mbox{GDR} \otimes %
\mbox{GQR}]_{1^-}$ and $[\mbox{GDR} \otimes 2^+_1]_{1^-}$ configurations
were not the only ones (among many different two-phonon $[1^- \otimes
2^+]_{1^-}$ states) to determine the total cross section, two-phonon $1^-$
configurations made of phonons of other multipolarities were omitted in
calculation.

While dealing with electromagnetic, or with Coulomb excitation from a $0^+$ ground
state, the priority attention has to be paid to the final states with the
total angular momentum and parity $J^{\pi} = 1^-$. Making use of the formalism
presented in ref. \cite{pon92} we have calculated firstly the cross section
of photoexcitation of two-phonon states $[\lambda_1^{\pi_1} \otimes
\lambda_2^{\pi_2}]_{1^-}$, where $\lambda_1^{\pi_1}$ and $\lambda_2^{\pi_2}$
are both natural $\lambda^{\pi^n}$ ($\pi^n = (-1)^{\lambda}$) and unnatural $%
\lambda^{\pi^u}$ ($\pi^u = (-1)^{\lambda + 1}$) parity phonons with
multipolarity $\lambda$ from 0 to 9. A phonon basis is obtained by solving
quasiparticle-RPA equations for each multipolarity $\lambda^{\pi}$ within
the quasiparticle-phonon model (QPM) \cite{QPM}. These equations provide a
set of one-phonon states $\lambda^{\pi}(i)$ with the same spin and parity,
but with different excitation energies $E(i)$ and internal fermion structure
of phonons; the index $i$ is introduced to distinguish between them. 
%The calculations were performed for nuclei $^{136}$Xe and $^{208}$Pb.

The results of the calculation for $^{136}$Xe and $^{208}$Pb integrated over
the energy interval from 20 to 35~MeV are presented in table~\ref{tab:1}.
Each configuration $[\lambda_1^{\pi_1} \otimes \lambda_2^{\pi_2}]$ in the
table means a sum over a plenty of two-phonon states made of phonons with a
given spin and parity $\lambda_1^{\pi_1}$, % and
$\lambda_2^{\pi_2}$, but different RPA root numbers $i_1$, %and
$i_2$ of its constituents 
%and thus, different excitation energy $E(i_1)+E(i_2)$:
\[
\sigma([\lambda_1^{\pi_1} \otimes \lambda_2^{\pi_2}]) = \sum_{i_1,i_2}
\sigma([\lambda_1^{\pi_1}(i_1) \otimes \lambda_2^{\pi_2}(i_2)])~. 
\]
The total number of two-phonon $1^-$ states included in this calculation for
each nucleus is about $10^5$. The absolute value of the photoexcitation of
any two-phonon state under consideration is negligibly small but altogether
they produce a sizable cross section. Table~\ref{tab:1} demonstrates that
different two-phonon configurations give comparable contributions to the
total cross section which decreases only for very high spins because of the
lower densities of such states. As a rule, unnatural parity phonons play
less important role than natural parity ones. For these reasons we presented
in the table only the sums for [natural$\otimes$unnatural] and [unnatural$%
\otimes$unnatural] two-phonon configurations.

The cross section of the photoexcitation of all two-phonon $1^-$ states in
the energy region 20-35~MeV from the ground state equals in our calculation
to 511~mb and 423~mb for $^{136}$Xe and $^{208}$Pb, respectively. It is not
surprising that we got a larger value for $^{136}$Xe than for $^{208}$Pb.
This is because the phonon states in Xe are composed of a larger number of
two-quasiparticle configurations due to the pairing. The same values for
two-phonon states with angular momentum and parity $J^{\pi} = 2^+$ are an
order of magnitude smaller. We point out that the direct excitation of $[1^-
\otimes 1^-]_{2^+}$ configuration or $[\mbox{GDR} \otimes \mbox{GDR}]_{2+}$
is negligibly weak. The calculated values should be compared to the cross
section of the photoexcitation of the single-phonon GDR which in our
calculation equals to 2006~mb and 2790~mb, respectively. A contribution of
two-phonon $1^-$ states to the total cross section at GDR energies is weaker
than at higher energies because of the lower density of two-phonon states
and lower excitation energy 
%to which photoexcitation cross section is linearly
%proportional for the case of E1-transitions.
and can be neglected considering the GDR itself.

For $^{208}$Pb photoexcitation cross sections are known from experimental
studies in $(\gamma, n)$ reactions up to the excitation energy about 25~MeV 
\cite{bel92,bel95}. It was shown that QPM provides a very good description
of the experimental data in the GDR region \cite{bel92}, while theoretical
calculations at higher excitation energies which account for contributions
from the single-phonon GDR and GQR$_{iv}$ essentially underestimated the
experimental cross section \cite{bel95}. The experimental cross sections
above 17~MeV are shown in Fig.~\ref{fig:1} together with theoretical
predictions. The results of the calculations are presented as strength
functions obtained with averaging parameter equal to 1~MeV. The contribution
to the total cross section of the GQR$_{iv}$ (short-dashed curve), the high
energy tail of GDR (long-dashed curve), and their sum (squared curve), are
taken from ref.~\cite{bel95}. The curve with triangles represents the
contribution of the direct excitation of the two-phonon states 
%in this energy region
from our present studies. The two-phonon states form practically a flat
background in the whole energy region under consideration. Summing together
the photoexcitation cross sections of all one- and two-phonon states we get
a solid curve which is in a very good agreement with the experimental data. 
%experiment.

From our investigation of photoexcitation cross sections we conclude that in
this reaction very many different two-phonon states above the GDR contribute
on a comparable level, forming altogether a flat physical background which
should be taken into account in the description of experimental data. On the
other hand, Coulomb excitation in relativistic heavy ion collisions provides
a unique opportunity to excite a very selected number of two-phonon states
by the absorption of two virtual  $\gamma$'s in a single process of
projectile-target interaction \cite{ber88}. Theoretically this process is
described using a the second order perturbation theory of the semi-classical
approach of A. Winther and K. Alder \cite{win79,ber88}. Since excitation cross
sections to second order are much weaker than to first order of the theory,
only two-phonon states connected to the ground states by two E1-transitions
can be observed. These two-phonon states have the structure $[1^-(i) \otimes
1^-(i^{\prime})]_{0^+,2^+}$ and form the DGDR.

To describe the properties of the DGDR in $^{208}$Pb %in the QPM framework
we applied the technique developed in refs.~\cite{pon96a,pon96b}. Within
this technique we couple two-phonon $[1^-(i) \otimes
1^-(i^{\prime})]_{0^+,2^+}$ states with many three-phonon ones. 
% to obtain the width of the DGDR.
Two-phonon states are built from the same six $1^-(i)$ RPA phonons as in
ref.~\cite{ber96} which have the largest B(E1) values and exhaust 90.6\% of
the EWSR. Only $0^+$ and $2^+$ components of the DGDR were considered (see
ref.~\cite{ber96} for the quenching of the $1^+$ component). We included in
the calculation about 7000 three-phonon states which have the largest matrix
elements for the interaction with the selected 21 two-phonon states.
Diagonalization of the QPM Hamiltonian on the basis of these two- and
three-phonon states yields a set of $0^+$ (and $2^+$) states; the wave
function of each state includes all two- and three-phonon terms with
different weights for different states. To distinguish between these states
we introduce the index $\nu$. 
%the maximum value of which equals to the sum of the selected two- and
%three-phonon components.
To obtain the Coulomb excitation cross section in second order perturbation
theory we also need to have the structure of the GDR as an intermediate
state. For that we calculated the GDR fine structure by coupling the same
six strongest one-phonon $1^-(i)$ states to about 1200 two-phonon $1^-$
states in the GDR region.

The cross section of the DGDR$(\nu_{0^+,2^+})$ states excitation via the GDR$%
(\nu_{1^-})$ states in this reaction equals to 
\begin{eqnarray}
\sigma_{\nu_{0^+,2^+}}
&=&    \big|      \sum_{\nu_{1^-}}
A(E_{\nu_{1^-}},E_{\nu_{0^+,2^+}})
<1^-(\nu_{1^-})||E1||0^+_{g.s.}>\nonumber \\
&\times&
<[1^- \otimes 1^-](\nu_{0^+,2^+})||E1||1^-(\nu_{1^-})>
\big|^2 \nonumber
\end{eqnarray}
where $A(E_1,E_2)$ is the energy dependent reaction amplitude. A
straightforward calculation of the two-step process for the excitation of
7000 DGDR states via 1200 intermediate GDR states is very time consuming.
Making use of a very smooth dependence of the function $A(E_1,E_2)$ on both
arguments we tabulated this function and used it in the final calculation of
the DGDR Coulomb excitation cross section in relativistic heavy ion
collisions. We considered the excitation of the DGDR in the projectile for a 
$^{208}$Pb (640 A$\cdot$MeV) $+^{208}$Pb collision, according to the
experiment in ref.~\cite{bor96}, and used the minimum value of the impact
parameter, b= 15.54~fm, corresponding to the parameterization of ref.~\cite
{ben89}.

The cross section for Coulomb excitation of the DGDR is presented in Fig.~%
\ref{fig:2} by the short-dashed curve as a strength function calculated with
an averaging parameter equal to 1~MeV. The width of the DGDR for $^{208}$Pb
is very close to $\sqrt{2}$ times the width of the single GDR, as for the
case of $^{136}$Xe \cite{pon96b}. The contribution of the background of the
two-phonon $1^-$ states to the total cross section is shown by a long-dashed
curve in the same figure. It was calculated in first order perturbation
theory. The role of the background in this reaction is much less important
than in photoexcitation studies. First, it is because in heavy ion
collisions we have a special mechanism to excite selected two-phonon states
in the two-step process. Second, the Coulomb excitation amplitude is
exponentially decreasing with the excitation energy, while the E1
photoexcitation amplitude is linearly increasing. Nonetheless, Fig.~\ref
{fig:2} shows that the direct excitation of two-phonon $1^-$ states cannot
be completely excluded from consideration of this reaction. Integrated over
the energy interval from 20 to 35~MeV these states give a cross section of
50.3~mb which should be compared to the experimental cross section in the
DGDR region for the $^{208}$Pb (640 A$\cdot$MeV) $+^{208}$Pb reaction which
is equal to 380~mb \cite{bor96}. Taking into account that the reported
experiment/theory enhancement is of about 30\% for this reaction, the two-phonon 
$1^-$ states omitted in extracting the DGDR strength from experimental cross
section maybe responsible for an appreciable part of the effect. Appreciable
values of one-step processes in DGDR excitation is not in contradiction with
experimental findings. It is known that Coulomb excitation of a projectile
in an $n$-step process has the following dependence on the target charge: $%
Z_T^{n \cdot (2 - \delta)}$. The reported value $n=1.8(3)$ in the DGDR
region for $^{208}$Pb \cite{bor96} allows for some contribution of one-step
transitions.

The solid line in Fig.~\ref{fig:2} is the sum of DGDR and two-phonon
background excitations in relativistic heavy ion collisions. As mentioned
above, this curve has a closer correspondence to the experimental cross
sections reported for the DGDR than the short-dashed one. Centroids and
widths calculated for the short-dashed and solid curves of this figure
display other interesting features. Taking into account the background of
two-phonon $1^-$ states results in a visual shift of the DGDR centroid by
-200~keV and a 16\% increase of the ``DGDR" width. The same effect, although
with large experimental uncertainties, was reported in an experiment with
respect to the harmonic picture of the DGDR excitation \cite{bor96}. We
point out that the effect of the visual shift of the DGDR centroid is even a
bit larger than anharmonicity effects examined for this nucleus in ref.~\cite
{lan97}.

The effect of the direct excitation of two-phonon states from the ground
state was considered in the present paper under the assumption that
these states do not interact between each other and with other one- and
three-phonon configurations.
The first type of interaction, unharmonicity effect, is rather weak as
discussed in the refs.~\cite{pon96b,lan97} and cannot change our
conclusions as far as the wide energy region is considered.
But one may argue that a weak coupling to one-phonon $1^-$ states with
much larger matrix elements of photoexcitation from the ground state may
reduce the calculated cross sections due to the destructive
interference with our two-phonon $1^-$ states.
Of course, the total contribution of this interference term integrated
from 0~MeV to infinity equals to zero because of orthogonality of wave
functions but it cannot be excluded from a consideration for a limited
energy interval.
To estimate its influence on our results we have performed a calculation
in which $1^-$ states in $^{208}$Pb where described by the wave function
consisting one- and two-phonon components.
The interaction between these components has been calculated
microscopically \cite{QPM}.
The full calculation with including all $10^5$ two-phonon $1^-$
configurations is not possible and we have done several tests
with different truncations of the two-phonon basis ending up in
total $10^3$ components for each test.
All one-phonon configurations with the energy below 35~MeV have been
included in calculations.
Selecting important two-phonon configurations we have chosen the ones
which have the largest matrix elements of the electromagnetic excitation
$<[\lambda_1^{\pi_1}(i_1) \otimes \lambda_2^{\pi_2}(i_2)]_{1^-} || E1 ||
g.s.>$ and
the largest matrix elements of interaction with one-phonon $1^-$
states.
The results of our test calculations are the following.

We do observe the effect of interference between transitions to one- and
two-phonon components of the wave functions of $1^-$ states.
In all runs the interference is constructive in the whole energy region
under consideration leading to an additional enhancement of the
calculated cross sections.
It becomes destructive only below 14.5~MeV reducing a little the total
E1-strength for the GDR.
In the DGDR energy region we estimate from our calculations the size of
the interference effect roughly as 30-50\% of what we get from the
direct excitation of non-interacting two-phonon $1^-$ states.
We think that this estimate is an upper limit for $^{208}$Pb because
many other two-phonon configurations will block somehow a coupling of
selected two-phonon configurations to one-phonon ones.
On the other hand, the interference between transitions to one- and
two-phonon configurations may play much important role for $^{136}$Xe
in which the interaction between these configurations is sufficiently
stronger as compared to $^{208}$Pb.
It requires an extra study.

In any case, the influence of a sea of two-phonon $1^-$ states above
20~MeV on calculated cross sections considered in the present paper in
the approximation of the pure two-phonon configurations can be taken as
a lower limit of the real effect.

In conclusion, we investigated the contribution of the direct excitation of
a sea of two-phonon states above the GDR to photo-neutron cross sections and
to cross sections in relativistic heavy ion Coulomb excitation. These states
form a flat structureless background which is very important for the correct
description of the photo-neutron data. Due to the peculiarities of Coulomb
excitation in heavy ion collisions, its role is less important in this case,
but its consideration appreciably removes the disagreement between
experiment and theory.

We thank Dr. V.V. Voronov for his suggestion to check the role of the
interference effect discussed at the end of the paper. This work was
partially supported by the RFBR (grant 95-02-05701) and the Brazilian
funding agencies CAPES, CNPq, FUJB/UFRJ, and by
MCT/FINEP/CNPQ(PRONEX) under contract No. 41.96.0886.00.

\begin{table}[th]
\caption{Cross sections for the direct photoexcitation of different
two-phonon configurations from the ground state integrated over the energy
interval from 20 to 35~MeV in $^{136}$Xe and $^{208}$Pb. The GDR cross
section integrated over the energy of its location is presented in the last
line for a comparison. }
\label{tab:1}
\begin{tabular}{rcc}
& \multicolumn{2}{c}{$\sigma_{\gamma}$, mb} \\ \cline{2-3}
Configuration & $^{136}$Xe & $^{208}$Pb \\ \hline
$[0^+ \otimes 1^-]_{1^-}$ & 4.4 & 3.9 \\ 
$[1^- \otimes 2^+]_{1^-}$ & 36.6 & 44.8 \\ 
$[2^+ \otimes 3^-]_{1^-}$ & 82.8 & 33.1 \\ 
$[3^- \otimes 4^+]_{1^-}$ & 101.0 & 56.7 \\ 
$[4^+ \otimes 5^-]_{1^-}$ & 68.9 & 37.3 \\ 
$[5^- \otimes 6^+]_{1^-}$ & 49.2 & 46.2 \\ 
$[6^+ \otimes 7^-]_{1^-}$ & 31.9 & 49.8 \\ 
$[7^- \otimes 8^+]_{1^-}$ & 13.6 & 12.5 \\ 
$[8^+ \otimes 9^-]_{1^-}$ & 4.9 & 9.0 \\ \hline
$
\begin{array}{c}
^9 \\[-3.5mm] 
\sum \\[-2.5mm] 
_{ \lambda_1,\lambda_2 = 1}
\end{array}
\hspace*{-2mm} [\lambda_1^{\pi_1^n} \otimes \lambda_2^{\pi_2^u}]_{1^-}$ & 
71.4 & 58.5 \\ 
$
\begin{array}{c}
^9 \\[-3.5mm] 
\sum \\[-2.5mm] 
_{ \lambda_1,\lambda_2 = 1}
\end{array}
\hspace*{-2mm} [\lambda_1^{\pi_1^u} \otimes \lambda_2^{\pi_2^u}]_{1^-}$ & 
46.7 & 71.1 \\ \hline
$
\begin{array}{c}
^9 \\[-3.5mm] 
\sum \\[-2.5mm] 
_{ \lambda_1,\lambda_2 = 0}
\end{array}
\hspace*{-2mm} [\lambda_1^{\pi_1^{n,u}} \otimes
\lambda_2^{\pi_2^{n,u}}]_{1^-}$ & 511.4 & 422.9 \\ \hline\hline
$[\mbox{GDR} \otimes \mbox{GDR}]_{2^+}$ & 0.33 & 0.22 \\ 
$
\begin{array}{c}
^9 \\[-3.5mm] 
\sum \\[-2.5mm] 
_{ \lambda_1,\lambda_2 = 1}
\end{array}
\hspace*{-2mm} [\lambda_1^{\pi_1^n} \otimes \lambda_2^{\pi_2^n}]_{2^+}$ & 
38.1 & 21.7 \\ \hline\hline
GDR\hspace*{10mm} & 2006 & 2790
\end{tabular}
\end{table}

\begin{figure}[tbp]
\caption{Photo-neutron cross section for $^{208}$Pb. Experimental data (dots
with experimental errors) are from ref.~\protect\cite{bel95}. The
long-dashed curve is the high energy tail of the GDR, the short-dashed curve
is the GQR$_{iv}$ and the curve with squares is their sum. The contribution
of two-phonon states is plotted by a curve with triangles. The solid curve
is the total calculated cross section. }
\label{fig:1}
\end{figure}

\begin{figure}[tbp]
\caption{The contribution for the excitation of two-phonon $1^-$ states
(long-dashed curve) in first order perturbation theory, and for two-phonon $%
0^+$ and $2^+$ DGDR states in second order (short-dashed curve). The total
cross section (for $^{208}$Pb (640 A$\cdot$MeV) $+^{208}$Pb) is shown by the
solid curve. }
\label{fig:2}
\end{figure}

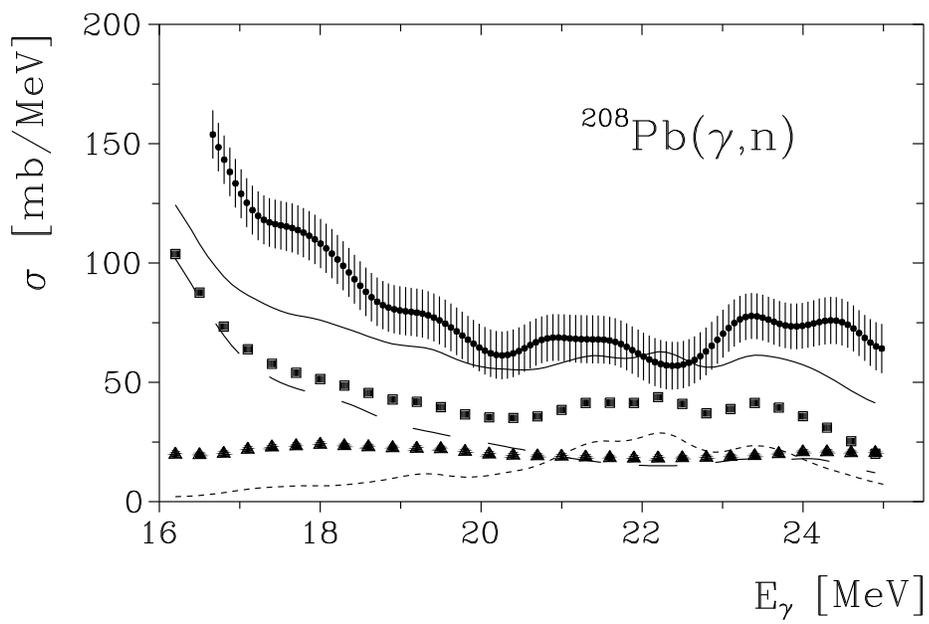
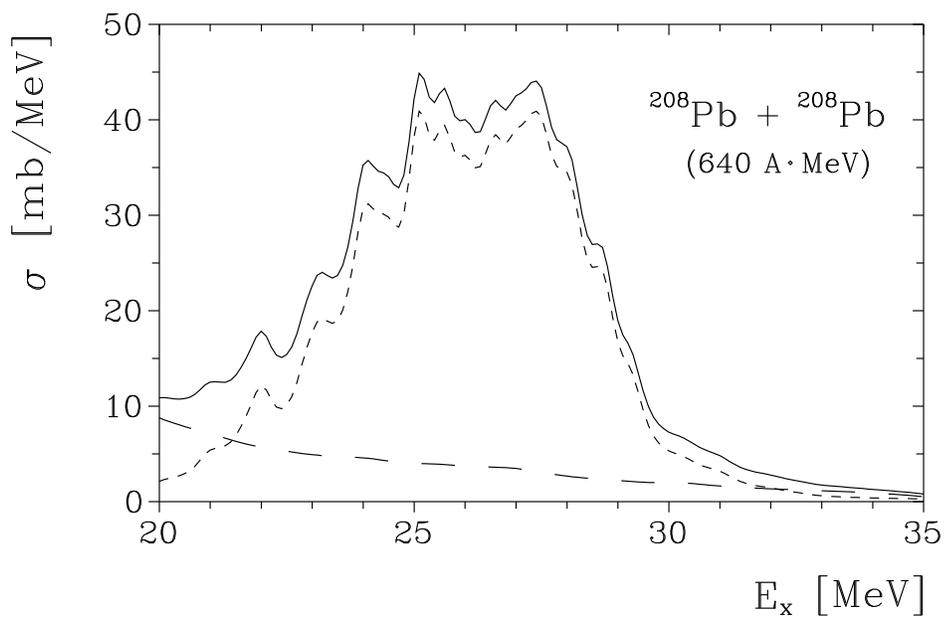
\end{document}